\documentclass[%
reprint, 
showkeys,
twocolumn,
superscriptaddress,
 amsmath,amssymb,
 aps, physrev,
floatfix,
]{revtex4-2}

\usepackage{graphicx}
\usepackage{dcolumn}
\usepackage{bm}
\usepackage{xcolor}
\usepackage[T1]{fontenc}
\usepackage[utf8]{inputenc}
\usepackage{textgreek}
\DeclareUnicodeCharacter{03B4}{\textdelta}
\DeclareUnicodeCharacter{2212}{\textminus}
\usepackage{placeins}

\begin{document}

\preprint{APS/123-QED}

\title{\textbf{Complete Suppression of Thermomagnetic Instabilities in Nb Superconducting Films by Combined Metallic Layers and Ion Irradiation} 
}%

\author{D. Carmo}
\affiliation{Departamento de F\'{i}sica, Universidade Federal 
de S\~{a}o Carlos, 13565-905, S\~{a}o Carlos, SP, Brazil}
\affiliation{Laborat\'{o}rio Nacional de Luz S\'{i}ncrotron, Centro Nacional de Pesquisa em Energia e Materiais, 13083-100, Campinas, SP, Brazil}
 \email{Contact author: fcolauto@df.ufscar.br}

\author{A. M. H. de Andrade}
\affiliation{Instituto de F\'{i}sica, Universidade Federal do 
Rio Grande do Sul, 91501-970, Porto Alegre, RS, Brazil}

\author{R. Giulian}
\affiliation{Instituto de F\'{i}sica, Universidade Federal do 
Rio Grande do Sul, 91501-970, Porto Alegre, RS, Brazil}

\author{A.~A.~ M.~Oliveira}
\affiliation{Instituto Federal de Educa\c{c}\~{a}o, Ci\^{e}ncia e Tecnologia 
de S\~{a}o Paulo, 13565-905, S\~{a}o Carlos, SP, Brazil}

\author{W. A. Ortiz}
\affiliation{Departamento de F\'{i}sica, Universidade Federal 
de S\~{a}o Carlos, 13565-905, S\~{a}o Carlos, SP, Brazil}

\author{F. Colauto}
\affiliation{Departamento de F\'{i}sica, Universidade Federal 
de S\~{a}o Carlos, 13565-905, S\~{a}o Carlos, SP, Brazil}

\date{\today}

\begin{abstract}
Thermomagnetic instabilities in superconducting films can trigger flux avalanches that disrupt the critical state and impair the performance of superconducting devices. Here we investigate the stabilization of Nb thin films subjected to a perpendicular magnetic field by two complementary approaches: the addition of normal-metal overlayers and Ar ion irradiation. Magnetization measurements and magneto-optical imaging show that the metallic layers suppress the onset of avalanches at low applied fields, whereas ion irradiation reduces the high-field portion of the instability region by shifting the upper threshold for avalanche activity to lower fields. When combined in the same sample, these two partial stabilization effects lead to complete suppression of thermomagnetic instabilities. In particular, a Nb film coated with a 1~\(\mu\)m-thick Cu layer after irradiation at a fluence of \(5\times10^{16}\)~ions/cm\(^2\) exhibits smooth magnetization curves and avalanche-free flux penetration. This combined treatment restores stable critical-state behavior and provides a practical route for stabilizing Nb superconducting films in thin-film superconducting technologies.
\end{abstract}


\maketitle


\section{Introduction}

Quantum computers can be based on superconducting qubits built from Josephson-junction circuits fabricated from superconducting thin films~\cite{clarke_superconducting_2008}.
Among the materials used in superconducting electronics, niobium (Nb) thin films are particularly attractive because of their relatively high superconducting transition temperature~\cite{gubin_dependence_2005}, low microwave losses~\cite{mcrae_materials_2020,kowsari_fabrication_2021}, and compatibility with standard microfabrication techniques as well as multilayer circuit architectures~\cite{chattaraj_materials_2026}. 
These properties make Nb films suitable for the fabrication of microwave resonators~\cite{wallraff_strong_2004}, transmission lines~\cite{tuckerman_flexible_2016}, SQUIDs~\cite{clarke_squid_2004}, and other elements of superconducting qubit circuits~\cite{oliver_materials_2013}. 
Although many modern transmon qubits employ Al-based Josephson junctions~\cite{de_leon_materials_2021}, niobium remains widely used in other parts of the superconducting circuitry, including resonators, ground planes, and wiring~\cite{siddiqi_engineering_2021,altoe_localization_2022,bruckmoser_niobium_2026}.

For these applications, however, the stability of the superconducting state under magnetic flux penetration becomes a critical issue, since thermomagnetic stability is crucial for the practical use of superconductors~\cite{mints_critical_1981}.
In this context, the occurrence of flux avalanches in superconducting films represents a significant limiting factor~\cite{vestgarden_nucleation_2018}. 
These events may be accompanied by transient electrical signals~\cite{mikheenko_nanosecond_2013}, induce local temperature rises and fluctuations~\cite{vestgarden_lightning_2012,jing_influences_2016}, alter the shielding-current distribution~\cite{vestgarden_nonlocal_2013}, and involve rapid, uncontrolled vortex rearrangement and flux motion~\cite{field_superconducting_1995,esquinazi_vortex_1999,bolz_dynamics_2003,welling_huge_2004,altshuler_colloquium_2004,aranson_dendritic_2005,baziljevich_dendritic_2014}. 
Therefore, understanding~\cite{joshi_quasiparticle_2023} and controlling~\cite{colauto_limiting_2013,carmo_trapping_2015,colauto_controlling_2021} flux avalanches in Nb thin films is essential for improving the reliability of superconducting devices and quantum circuits.

Several strategies have been proposed to suppress thermomagnetic instabilities in superconducting films. 
One of the most effective is to cover the superconductor with a normal metallic layer, provided that the layer is sufficiently uniform and placed in direct contact with, or in close proximity to, the surface of the superconducting film~\cite{colauto_suppression_2010}. 
When first reported, this effect was attributed to the ability of the metal layer to act as a heat sink, efficiently dissipating into the environment the heat generated during flux avalanches~\cite{baziljevich_origin_2002}. 
Subsequent experiments demonstrated, however, that suppression may occur even without thermal contact, indicating that the dominant mechanism is electromagnetic braking produced by eddy currents induced in the adjacent normal metal layer, which oppose rapid flux redistribution~\cite{colauto_suppression_2010}. 
Later theoretical studies reconciled these interpretations by showing that both the thermal and electrodynamic properties of the metal layer contribute to stabilization through distinct mechanisms: enhanced thermal conductivity suppresses the nucleation of the thermomagnetic instability, while increased electrical conductivity introduces inductive braking that limits avalanche propagation and branching~\cite{vestgarden_thermomagnetic_2013,vestgarden_inductive_2014}.

Another possible route to modify the behavior of superconducting films, which we explored in this work, is ion irradiation. This process can create structural defects that act as vortex-pinning centers~\cite{civale_irradiation-enhanced_1992,hua_vortex_2010,smylie_effect_2016,tran_local_2026}. Under suitable irradiation conditions, the resulting defect landscape can enhance the critical current density, which is beneficial for applications requiring large currents and high magnetic fields~\cite{massee_imaging_2015,budhani_effects_1993,behler_vortex_1994,li_enhancement_2026}. 
In addition, this treatment can induce structural modifications in superconducting films after deposition~\cite{sahoo_structural_2015,ozaki_route_2016}. 
By selecting appropriate parameters, it is possible, for example, to promote stress relaxation in the film~\cite{pease_modifications_1984,daghero_effect_2017}, which may otherwise develop during the deposition process~\cite{hoffman_internal_1977,booi_intrinsic_1993}. 
Because this treatment modifies the physical properties of superconducting films, a central question addressed here is how it affects thermomagnetic instabilities.

In this work, we investigate the stabilization of superconducting Nb films using two complementary approaches: (i) covering the films with layers of different normal metals and (ii) irradiating the films with argon ions at different fluences. 
The combination of these modifications within the same sample yields a hybrid system free of thermomagnetic instabilities, thereby restoring the ability of the superconductor to shield magnetic fields and sustain supercurrents.

\section{Materials and Methods}

Nb films were deposited on 4-inch-diameter Si(100) substrates by dc magnetron sputtering in an AJA Orion-8 UHV chamber with a base pressure below $2\times10^{-8}$~Torr. Square samples were subsequently defined by optical lithography. Samples with a lateral size of 2.0 mm have a thickness of 180 nm, whereas those with a lateral size of 2.5 mm have a thickness of 165 nm. 
The film thicknesses were determined by x-ray reflectivity (XRR) using an X'Pert Pro MRD XL PANalytical diffractometer with Cu K$\alpha$ radiation ($\lambda = 1.5418$~\AA), with an estimated uncertainty below 1

The first set of square samples was used to fabricate superconductor/normal-metal hybrid structures by depositing metallic overlayers on Nb samples with a lateral size of 2.0~mm, extending beyond their edges. Three samples were prepared with Cu, Ag, and Au overlayers. Prior to metal deposition, the Nb films were exposed to air, resulting in the natural formation of an approximately 5-nm-thick Nb$_2$O$_5$ oxide layer~\cite{sokhey_oxidation_2010}. The metallic overlayers were deposited by dc magnetron sputtering using the same system employed for Nb growth. Their thickness, as determined by scanning electron microscopy, was \((1.0 \pm 0.1)~\mu\mathrm{m}\). Overlayers of this thickness have previously been shown to effectively suppress thermomagnetic instabilities~\cite{choi_enhancement_2004}.

The second batch of square samples was used for Ar$^{+}$ irradiation at room temperature with an ion energy of 350~keV at normal incidence. 
At this energy, the ions fully traverse the Nb film and are implanted into the substrate. 
The ion fluence was controlled by the irradiation time and ranged from $1\times10^{15}$ to $1\times10^{17}$~ions/cm$^{2}$. 
During irradiation, the beam current density was kept constant at 0.5~$\mu$A/cm$^{2}$ in order to avoid excessive heating of the films. 
The fluence values reported for the irradiated samples are nominal values, estimated from the beam current density and irradiation time, with an uncertainty below 5\%.

For clarity, the samples are identified using a simple nomenclature that indicates the presence of metallic coatings or ion irradiation. 
For example, Nb/Cu denotes Nb films coated with Cu, whereas Nb-I1E16 refers to Nb films irradiated with Ar$^{+}$ at a fluence of $1\times10^{16}$ ions/cm$^{2}$. 
The suffix 2.0 or 2.5 indicates the lateral size of the square sample in millimeters. 
The investigated samples are summarized in Table~\ref{Tab-sampleID}.

\begin{table}[h]
\centering
\caption{Identification and main characteristics of the investigated Nb films, including metallic coatings and ion-irradiated samples.}
\begin{tabular}{ll}
\hline
Sample & Feature \\
\hline
Nb-2.0 & plain film \\
Nb/Ag-2.0 & Ag layer \\
Nb/Au-2.0 & Au layer \\
Nb/Cu-2.0 & Cu layer \\

Nb-2.5 & plain film \\
Nb-I1E15-2.5 & $1\times10^{15}$ ions/cm$^{2}$ \\
Nb-I5E15-2.5 & $5\times10^{15}$ ions/cm$^{2}$ \\
Nb-I1E16-2.5 & $1\times10^{16}$ ions/cm$^{2}$ \\
Nb-I5E16-2.5 & $5\times10^{16}$ ions/cm$^{2}$ \\
Nb-I1E17-2.5 & $1\times10^{17}$ ions/cm$^{2}$ \\

Nb/Cu-2.5 & Cu layer \\
Nb/Cu-I1E15-2.5 & Cu layer + $1\times10^{15}$ ions/cm$^{2}$ \\
Nb/Cu-I5E15-2.5 & Cu layer + $5\times10^{15}$ ions/cm$^{2}$ \\
Nb/Cu-I1E16-2.5 & Cu layer + $1\times10^{16}$ ions/cm$^{2}$ \\
Nb/Cu-I5E16-2.5 & Cu layer + $5\times10^{16}$ ions/cm$^{2}$ \\
Nb/Cu-I1E17-2.5 & Cu layer + $1\times10^{17}$ ions/cm$^{2}$ \\
\hline
\end{tabular}
\label{Tab-sampleID}
\end{table}

Electrical resistivity measurements were performed in a Quantum Design PPMS-6000 using the van der Pauw method~\cite{van_der_pauw_method_1958}. 
Electrical contacts between the samples and the puck sample holder were made using a TPT HB05 wire bonder. 
Magnetic measurements were carried out in a Quantum Design MPMS-5S SQUID magnetometer, which allows both ac susceptibility and dc magnetization measurements. The superconducting transition temperature $T_c$ was determined from the real component of the ac magnetic moment, while dc magnetization measurements were used to investigate thermomagnetic instabilities.

Magneto-optical imaging (MOI) was performed using a standard setup described in Refs.~\cite{vlasko-vlasov_magneto-optical_1999,Polyanskii_magneto-optical_2004}. 
A Bi-substituted ferrite garnet film with in-plane magnetization was placed directly on top of the sample and served as a Faraday-active indicator of the local magnetic flux density. 
The experiments were carried out in an Oxford MicrostatHe-R optical cryostat equipped with an optical window for imaging, while a magnetic field perpendicular to the sample surface was applied using a pair of split coils. 
Images were acquired with an Olympus BX-LRA-2 microscope fitted with a U-PO3 polarizer, a U-AN360-3 analyzer, and a 5$\times$ MPlanFL N objective, using a Retiga 4000R QImaging CCD camera.
In this configuration, the image brightness is proportional to the local perpendicular magnetic flux density.

Scanning electron microscopy (SEM) measurements were performed using a FEI Magellan 400 L field-emission gun microscope. SEM was used to determine the thickness of the metallic overlayers and to examine the surface morphology of the samples, including film continuity, surface homogeneity, and irradiation-induced changes in the microtexture of the Nb films.

\section{Sample Characterization}

\begin{figure}[t!]
\centering\includegraphics[width=1.0\linewidth]{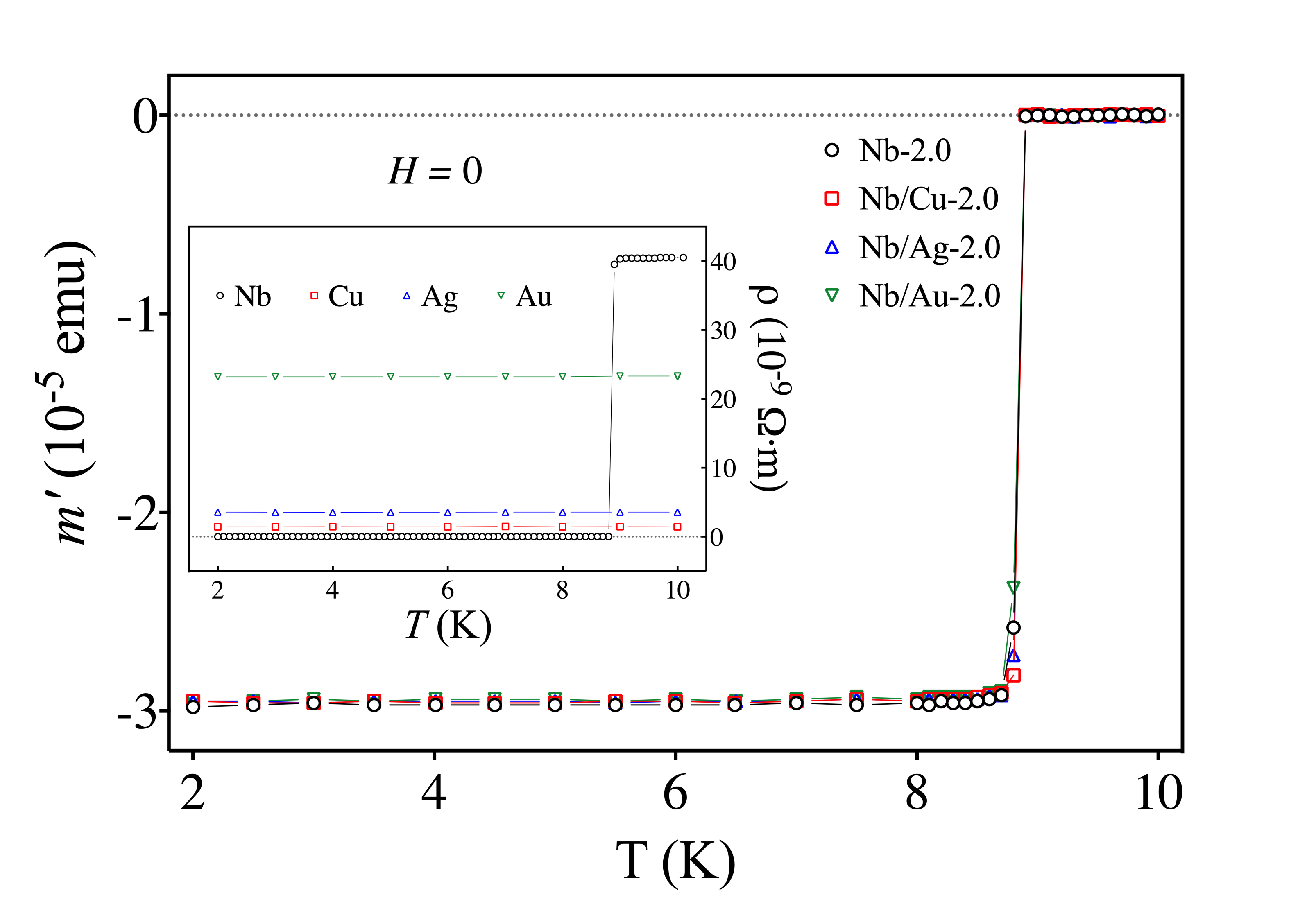}
\caption{
Temperature dependence of the real part of the ac magnetic moment for the uncovered Nb-2.0 film and the corresponding samples with Cu, Ag, and Au overlayers. Inset: electrical resistivity as a function of temperature for the Nb film and for the metallic layers deposited on Si substrates.
}
\label{Fig1}
\end{figure}

The superconducting transition temperature, $T_c$, of the samples was determined from measurements of the real component of the ac magnetic moment, $m'$, at zero applied dc magnetic field ($H=0$).
The uncovered Nb-2.0 film exhibits $T_c = (8.8 \pm 0.2)$~K. Similar values of $T_c$ and transition width were found for all superconductor/normal-metal hybrid samples, as shown in the main panel of Fig.~\ref{Fig1}.

To determine the electrical resistivity of the metallic overlayers independently of the hybrid structures, Cu, Ag, and Au films were simultaneously deposited onto Si substrates under the same conditions used for the coated samples. This procedure enabled low-temperature resistivity measurements ($\rho$) of the normal metals without contribution from the superconducting Nb layer, as shown in the inset of Fig.~\ref{Fig1}. For these metals, $\rho$ remains essentially constant below 10~K, whereas the Nb film exhibits a sharp superconducting transition in zero applied field. The corresponding resistivity values, together with the residual resistivity of Nb, $\rho_0$, estimated at 10~K, are summarized in Table~\ref{Tab-camadas}.

\begin{figure}[t!]
\centering\includegraphics[width=1.0\linewidth]{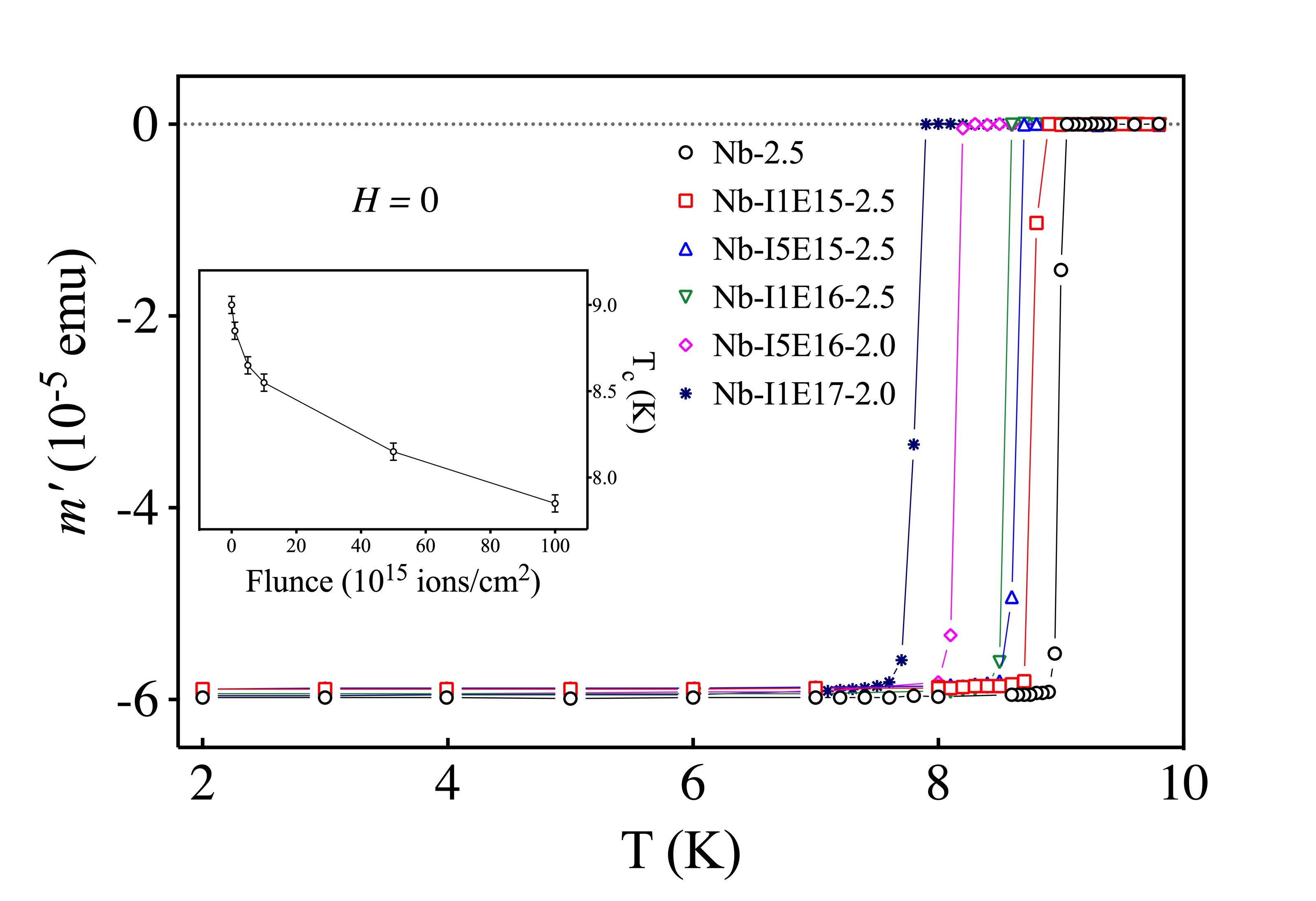}
\caption{
Temperature dependence of the real part of the ac magnetic moment for the pristine Nb-2.5 film and the corresponding Ar$^{+}$-irradiated samples at different fluences. Inset: superconducting transition temperature as a function of ion fluence, obtained from the main-panel data.
}
\label{Fig2}
\end{figure}

The main panel of Fig.~\ref{Fig2} shows the superconducting transitions of the irradiated samples, measured at $H=0$.
The pristine Nb-2.5 film exhibits $T_c = (9.0 \pm 0.1)$~K. As the ion fluence increases, $T_c$ shifts progressively to lower temperatures, as shown in the inset of Fig.~\ref{Fig2}. In contrast, the transition width remains nearly unchanged, and the irradiated films reach the same saturation diamagnetic signal as the pristine film.

\begin{table}[h]
\caption{Resistivity $\rho$, conductivity $\sigma$, and parameter $S$ (see main text) for the Nb film and the metallic layers. In the case of Nb, \(\rho\) and \(\sigma\) correspond to the residual values \(\rho_{0}\) and \(\sigma_{0}\), respectively.}
\begin{tabular}{cccc}
\hline
Metal & $\rho$ (n$\Omega$m) & $\sigma$ (MS/m) & $S$ \\
\hline
Nb & $40.4 \pm 0.3$ & $24.8 \pm 0.2$ & -- \\
Cu & $1.4 \pm 0.1$ & $710 \pm 50$ & $175 \pm 13$ \\
Ag & $3.5 \pm 0.1$ & $286 \pm 8$ & $70 \pm 2$ \\
Au & $23.2 \pm 0.4$ & $43.1 \pm 0.7$ & $10.6 \pm 0.2$ \\
\hline
\end{tabular}
\label{Tab-camadas}
\end{table}

\begin{figure}[t!]
\centering\includegraphics[width=1.0\linewidth]{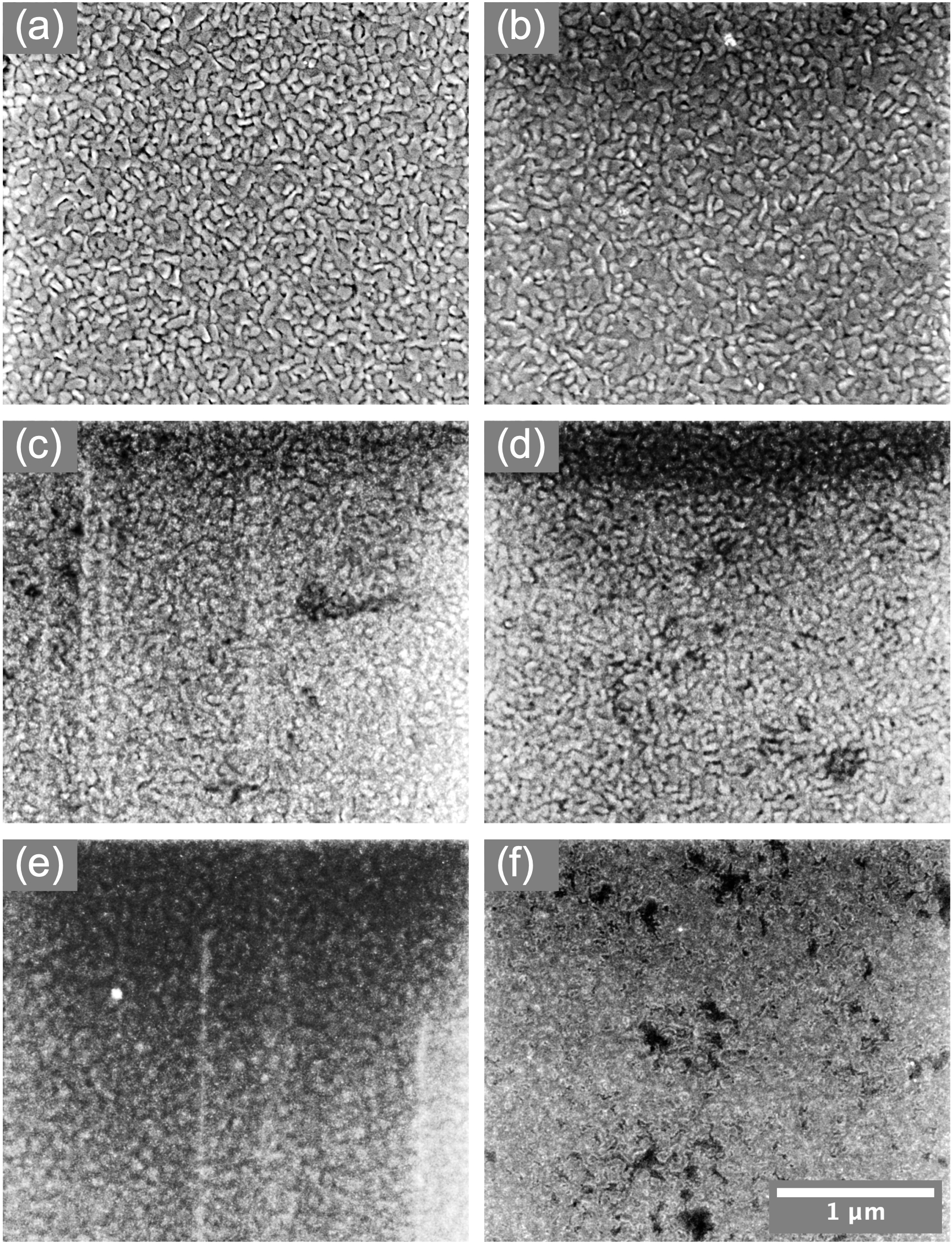}
\caption{
Top-view SEM images of the pristine Nb film and the corresponding Ar$^{+}$-irradiated films: (a) Nb-2.5, (b) Nb-I1E15-2.5, (c) Nb-I5E15-2.5, (d) Nb-I1E16-2.5, (e) Nb-I5E16-2.5, and (f) Nb-I1E17-2.5.
}
\label{Fig3}
\end{figure}

Figure~\ref{Fig3} shows top-view SEM images of the pristine and Ar$^{+}$-irradiated Nb-2.5 films. The pristine sample, Fig.~\ref{Fig3}(a), exhibits a continuous and homogeneous surface with a fine granular morphology, characteristic of a well-coalesced polycrystalline Nb film. The grains have a typical lateral size in the range of \(50\) to \(100\) nm. After irradiation at a fluence of \(1\times10^{15}\) ions/cm$^{2}$, Fig.~\ref{Fig3}(b), the surface morphology remains very similar to that of the pristine film, indicating that the continuity of the film is preserved and that only minor surface modifications are introduced at this fluence. As the fluence increases to \(5\times10^{15}\) and \(1\times10^{16}\) ions/cm$^{2}$, Figs.~\ref{Fig3}(c) and \ref{Fig3}(d), the surface texture becomes progressively smoother and the granular features become less distinct. For the highest fluences, \(5\times10^{16}\) and \(1\times10^{17}\) ions/cm$^{2}$, Figs.~\ref{Fig3}(e) and \ref{Fig3}(f), this trend becomes even more pronounced, with a substantial reduction in topographic contrast and a marked attenuation of the original granular texture. Nevertheless, no evidence of severe cracking, delamination, or substrate exposure is observed. Overall, the SEM images indicate that Ar$^{+}$ irradiation progressively alters the surface microtexture of the Nb films while preserving their global continuity.

\section{Results and Discussion}
\subsection{Metallic Layers}

The influence of a normal-metal overlayer on the stability of superconducting films is well established~\cite{baziljevich_origin_2002, choi_suppression_2009, colauto_suppression_2010, stahl_avalanche_2013}. 
The role of the metal properties is further supported by theoretical studies of thermomagnetic instabilities~\cite{aranson_dendritic_2005, denisov_dendritic_2006, denisov_onset_2006}. 
In particular, extensions of the standard framework to superconductor/metal bilayers predict that avalanche suppression becomes more effective as the product of the metal conductivity and thickness increases, in agreement with our experimental observations~\cite{vestgarden_thermomagnetic_2013, vestgarden_inductive_2014}.

The enhancement of thermomagnetic-stability by increasing the thickness of a normal-metal overlayer has also been demonstrated experimentally~\cite{choi_enhancement_2004}. 
In that study, Au layers with thicknesses up to 2.55~$\mu$m were deposited on MgB$_2$ films, establishing Au as an effective coating material because of its chemical stability and high electrical conductivity. 
Since the suppression mechanism also depends on the electrical and thermal properties of the adjacent normal metal, we extend this approach here to other metallic overlayers.

\begin{figure}[t!]
\centering\includegraphics[width=1.0\linewidth]{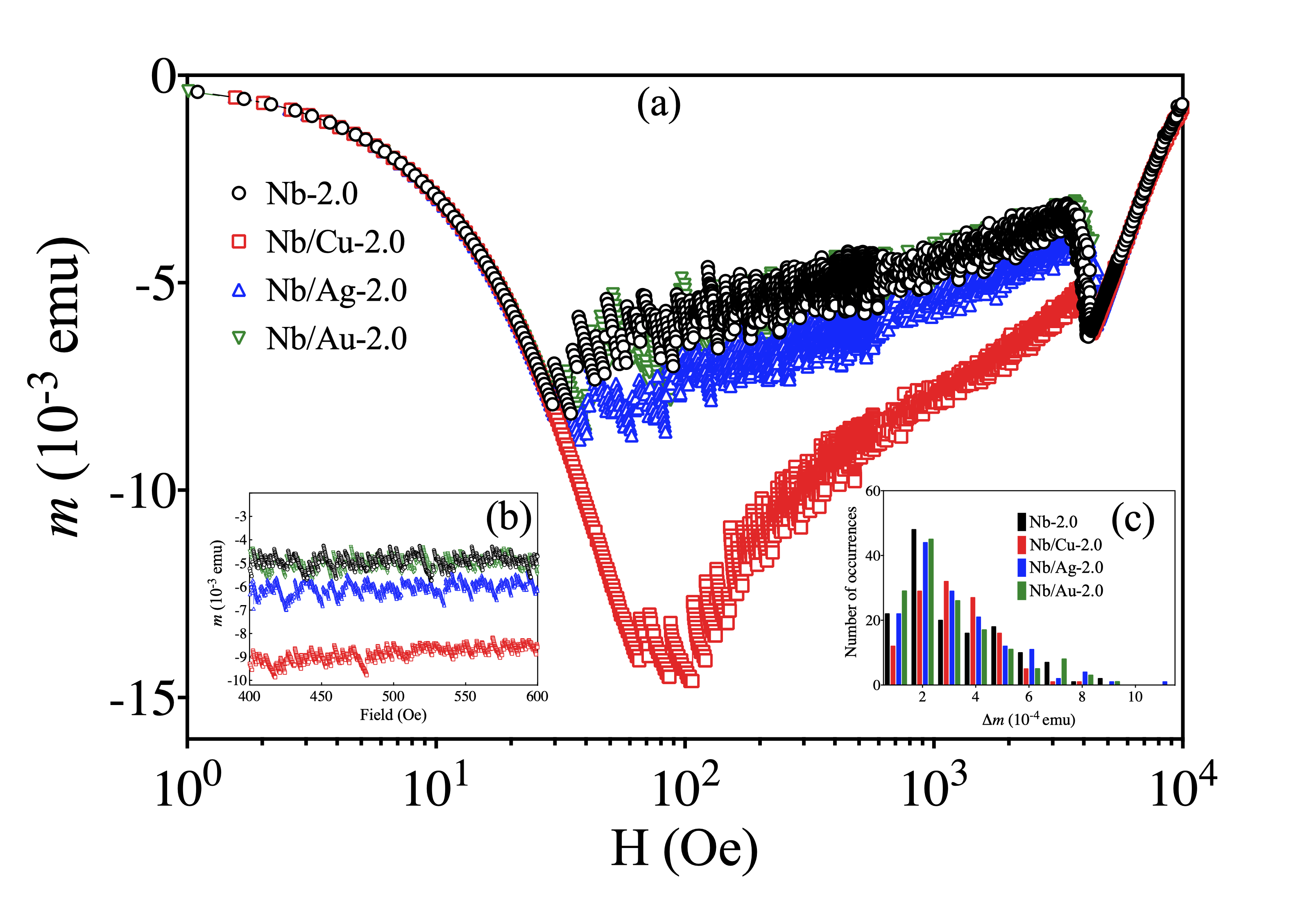}
\caption{
Magnetic moment as a function of applied magnetic field for the Nb-2.0 film and the corresponding samples coated with Cu, Ag, and Au, measured after zero-field cooling to 3.0~K. Panel (a) is plotted on a logarithmic field scale to accommodate the magnetization over a broad field range. Panel~(b) shows an enlarged view of the field interval used for the statistical analysis of the magnetization jumps shown in panel~(c).
}
\label{Fig4}
\end{figure}

Figure~\ref{Fig4} shows that the addition of a normal-metal layer enhances the diamagnetic response and leads to partial suppression of thermomagnetic instabilities, as evidenced by magnetic-moment measurements as a function of the applied field. To resolve the successive jumps in magnetic moment over a broad field range, the main panel is plotted on a logarithmic field scale, while the insets present an enlarged view of a selected region on a linear scale (b) and statistical analyses of the size, $\Delta m$, of the jumps. Before each measurement, the samples were zero-field cooled from above $T_c$ to 3.0~K. For comparison, the isotherm of Nb-2.0 is shown as black circles.

Both the virgin branches and the final portions of the curves, as the samples approach the normal-state transition, are smooth for all films, indicating flux penetration in the absence of avalanches. 
The overlap of the curves in these regions shows that the normal-metal layers do not degrade the superconductivity of the Nb films and further indicates that the samples exhibit similar diamagnetic responses there, consistent with their comparable superconducting properties and dimensions. 
By contrast, in the intermediate-field range the curves display a series of abrupt discontinuities associated with flux avalanches~\cite{colauto_boundaries_2008}. The field values defining this instability regime are identified as the lower and upper threshold fields. 
The Cu-coated sample remains in the smooth diamagnetic regime up to higher fields than the uncoated Nb film does, with a lower threshold field of about 65~Oe for Nb/Cu-2.0, compared with about 30~Oe for Nb-2.0. In addition, Nb/Cu-2.0 exhibits a substantially stronger diamagnetic response than Nb-2.0, indicating a correspondingly higher shielding-current density. 
The Ag-coated sample, Nb/Ag-2.0, also exhibits an enhanced diamagnetic response relative to Nb-2.0, whereas Nb/Au-2.0 shows only a slight increase and differs little from the uncoated film.

Because the metallic overlayers have comparable thicknesses, the differences in diamagnetic response can be directly associated with the transport properties of the adjacent normal metal. A comparison of the low-temperature resistivities in Fig.~\ref{Fig1} with the magnetization curves in Fig.~\ref{Fig4} shows that the suppression of thermomagnetic instabilities becomes more effective as the resistivity of the metallic layer decreases. This enhanced stabilization is accompanied by a stronger diamagnetic response, indicating a more complete recovery of the shielding-current state that is otherwise disrupted by flux avalanches.

\begin{figure}[t!]
\centering\includegraphics[width=1.0\linewidth]{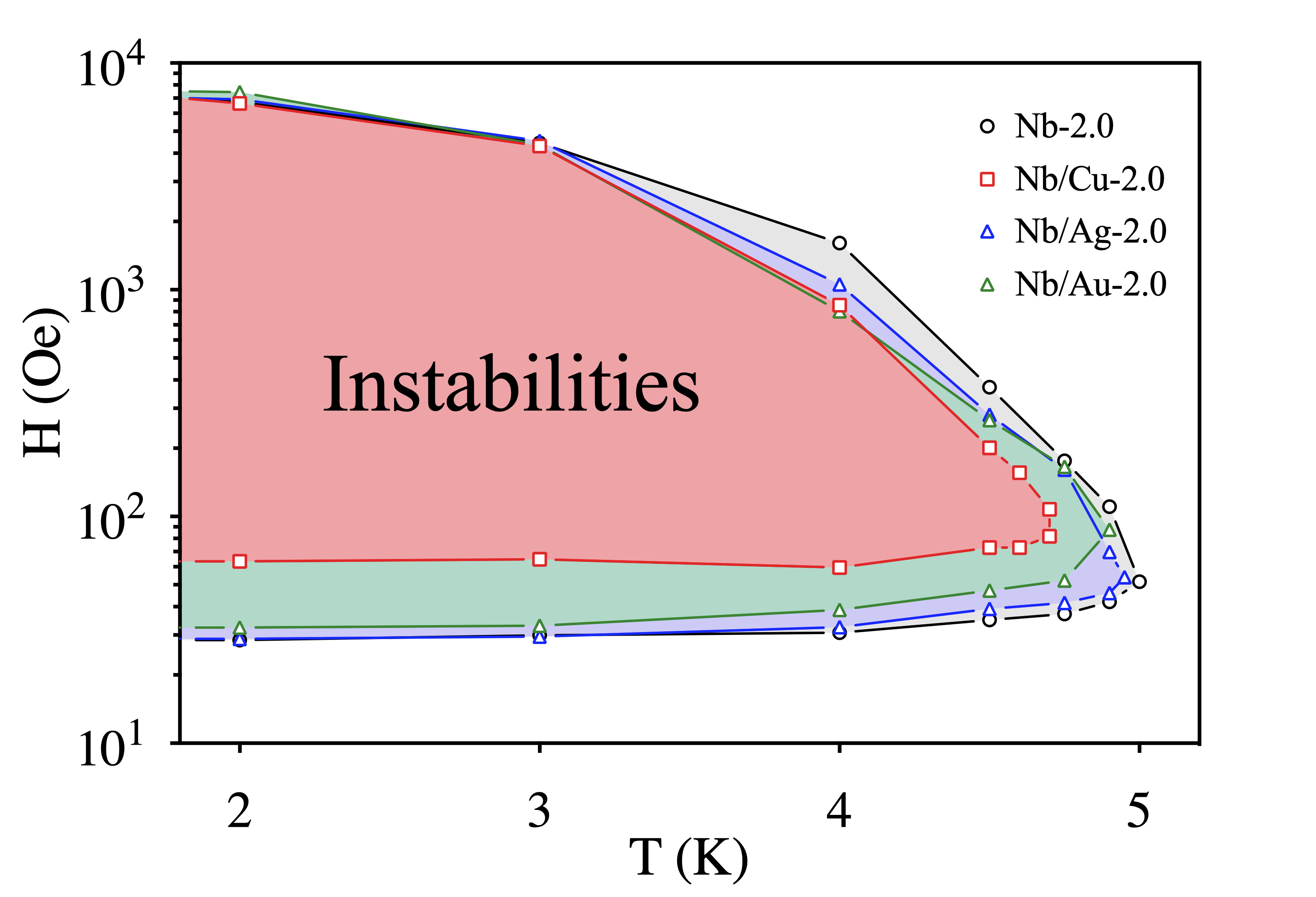}
\caption{
Field-temperature (\(H\)-\(T\)) diagram for the Nb-2.0 film and the corresponding samples coated with Cu, Ag, and Au. The applied magnetic field is shown on a logarithmic scale. The region enclosed by the lower and upper threshold fields defines the field and temperature range over which thermomagnetic instabilities occur.
}
\label{Fig5}
\end{figure}

The lower and upper threshold fields, identified from magnetization isotherms measured at different temperatures for the uncoated Nb film and the hybrid systems, are summarized in the \(H\)-\(T\) diagram of Fig.~\ref{Fig5}. The region bounded by the lower and upper threshold fields defines the thermomagnetic-instability regime, in which flux avalanches are observed~\cite{colauto_boundaries_2008, colauto_mapping_2007}. Outside this region, flux penetrates the superconductor smoothly, without intermittent avalanche activity.

Taking the boundary of the Nb-2.0 as a reference, the presence of a normal-metal overlayer is seen to reduce the region of the \(H\)-\(T\) diagram over which thermomagnetic instabilities occur.
This reduction is most pronounced for Nb/Cu-2.0, whose threshold boundaries enclose the smallest instability region among the investigated samples, indicating that Cu provides the most effective suppression within the field and temperature range explored here.
For the other metallic coatings, however, the extent of the instability region alone is not sufficient to establish a definitive ranking of their stabilizing efficiency.
Although Nb/Au-2.0 appears to define a smaller instability region than Nb/Ag-2.0, these boundaries reflect only the onset and termination of the instability regime during isothermal field increase.
A more complete assessment of the effectiveness of each normal-metal layer therefore requires consideration of additional quantities, such as the magnitude of the diamagnetic response, the amplitude of the jumps in the magnetization isotherms, and the morphology of the avalanches~\cite{zadorosny_morphology_2013}.

A thermomagnetic model for hybrid superconductor/normal-metal systems introduces the dimensionless suppression parameter \(S\)~\cite{vestgarden_inductive_2014},
\begin{equation}
S \equiv \frac{d_m \sigma_m}{d_s \sigma_{sn}},
\label{eq:S}
\end{equation}
where \(d_m\) and \(d_s\) are the thicknesses of the normal-metal layer and the superconducting film, respectively, and \(\sigma_m\) and \(\sigma_{sn}\) are the electrical conductivities of the metal and the superconductor in the normal state. For \(S \gg 1\), thermomagnetic instabilities are expected to be strongly suppressed. The values of \(S\) for the hybrid structures studied here are listed in Table~I, together with the electrical conductivity, \(\sigma = 1/\rho\), of the metallic layers and the Nb film. This parameter predicts more effective suppression as the conductivity and thickness of the normal-metal layer increase. Consistent with the experimental observations, \(S\) is largest for Nb/Cu, followed by Nb/Ag and Nb/Au.

\begin{figure}[t!]
\centering\includegraphics[width=1.0\linewidth]{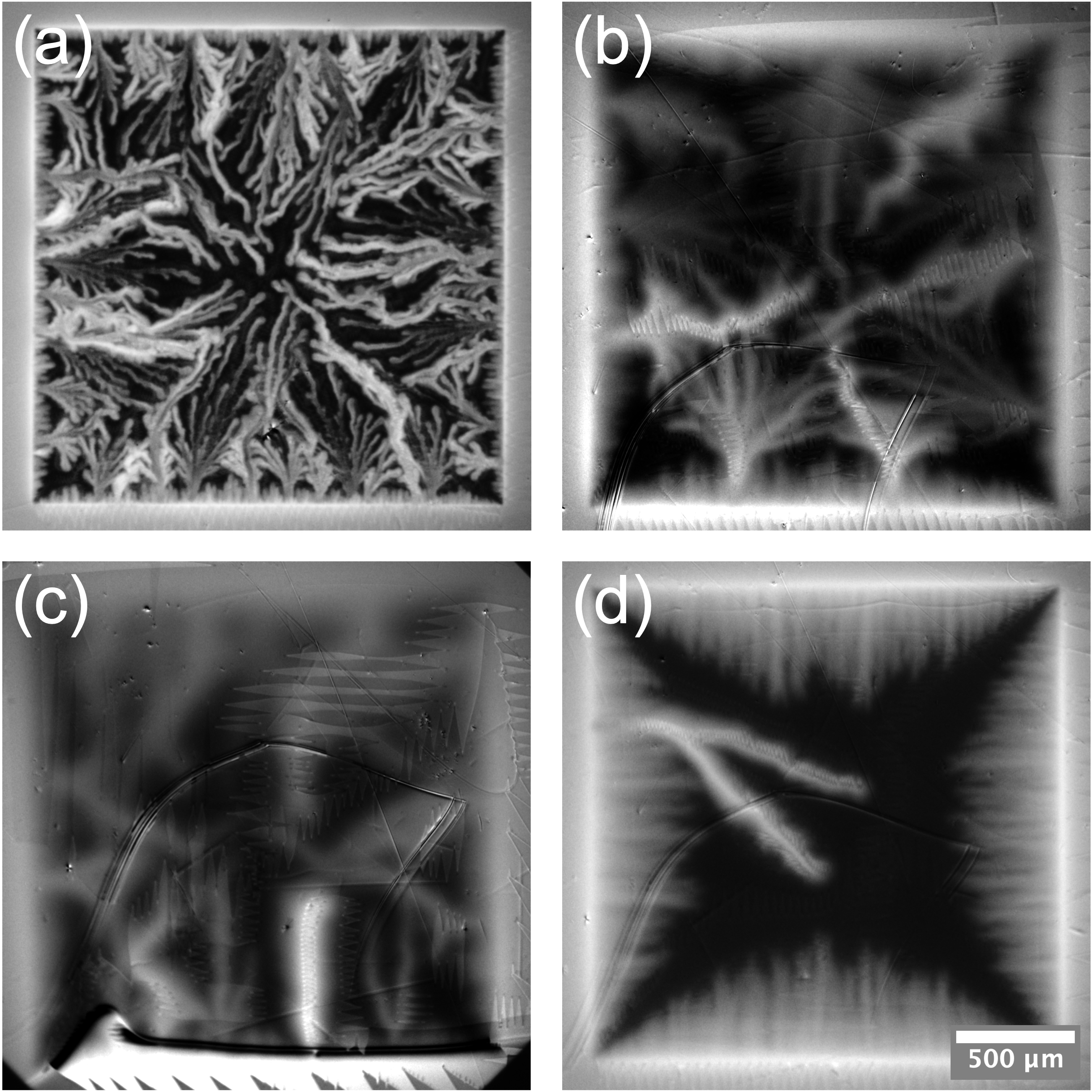}
\caption{
Magneto-optical images of flux penetration in the uncoated Nb film and the corresponding samples coated with metallic layers, acquired at 3.0~K and 100~Oe after zero-field cooling. In panel (a), the Nb-2.0 film exhibits numerous highly branched dendritic avalanches occupying most of the sample area. In panel (b), the avalanches in Nb/Au-2.0 are shorter and wider. In panel (c), the avalanches in Nb/Ag-2.0 are less branched and display a fingerlike morphology. In panel (d), avalanche activity in Nb/Cu-2.0 is strongly reduced, and flux penetration is predominantly smooth, consistent with critical-state behavior.
}
\label{Fig6}
\end{figure}

Figure~\ref{Fig6} presents magneto-optical images of the flux-avalanche morphology. 
To enable direct comparison with the magnetization measurements, the images were acquired under the same zero-field-cooling and stepwise field-increase conditions, at 3.0~K and an applied field of 100~Oe. 
In panel (a), corresponding to Nb-2.0, the avalanches are thin, long, and numerous, characteristic of a highly unstable regime.
In panel (b), for Nb/Au-2.0, they become shorter and wider, indicating partial suppression of the branching dynamics. 
In panel (c), for Nb/Ag-2.0, only a few avalanches are observed, with a morphology closer to flux fingers and a lower occurrence frequency. 
In panel (d), for Nb/Cu-2.0, only a relatively large bifurcated avalanche is observed.

The progressive reduction in the number, branching, and spatial extent of the avalanches is consistent with the enhanced stabilization provided by the metallic layers and is directly reflected in the magnetization measurements~\cite{colauto_mapping_2007}. Around 100~Oe in Fig.~\ref{Fig4}, Nb-2.0 exhibits numerous fluctuations, whereas such events become less frequent in the coated samples. For Nb/Cu-2.0, the few but large magnetization jumps indicate less frequent avalanches with larger spatial extent, in agreement with Fig.~\ref{Fig6}(d).

The magneto-optical images of the metal-coated samples appear more diffuse because the overlayer increases the distance between the superconducting surface and the indicator film. Even so, the suppression of avalanches remains evident. In terms of the suppression parameter \(S\), the extent of avalanche branching decreases as \(S\) increases. In addition, the hybrid samples exhibit a smaller total number of avalanches than the uncoated Nb film. Another indication of the stabilization induced by the adjacent metallic layer is the smooth flux penetration from the sample edges, characteristic of critical-state behavior. Taken together, the isothermal magnetization measurements and the MOI images provide consistent evidence that the normal-metal layers have a stabilizing effect on the superconducting film.

In Fig.~\ref{Fig4}, inset (b) shows the distribution of jump occurrences as a function of the magnetic-moment jump amplitude, \(\Delta m\), expressed in emu. The analysis was performed over the 400--600~Oe interval of the 3.0~K magnetization isotherms shown in Fig.~\ref{Fig4}(c), for which the measurements were carried out with a field step of 0.2~Oe. Within this approach, \(\Delta m\) is taken as a measure of the flux-avalanche size~\cite{colauto_mapping_2007}.
The Nb-2.0 sample exhibits a maximum in the interval \(1\times10^{-4} < \Delta m < 2\times10^{-4}\), indicating that most instabilities occur as relatively low-amplitude events. By contrast, for Nb/Cu-2.0 the maximum shifts to the interval \(2\times10^{-4} < \Delta m < 3\times10^{-4}\), and the distribution becomes noticeably broader. This behavior indicates that, although the total number of discontinuities is reduced, the remaining events occur with larger amplitudes and therefore correspond to larger flux avalanches.

Nb/Ag-2.0 and Nb/Au-2.0 exhibit similar tendencies, although less pronounced. In these samples, the number of events decreases moderately and the distribution shifts only slightly toward higher \(\Delta m\) values, without the clear displacement observed for Nb/Cu-2.0. Overall, these results show that the modification of the instability-amplitude distribution depends on the adjacent metallic layer and is significantly more pronounced for Cu than for Ag or Au.

\subsection{Ion Irradiation}

\begin{figure}[b]
\centering\includegraphics[width=1.0\linewidth]{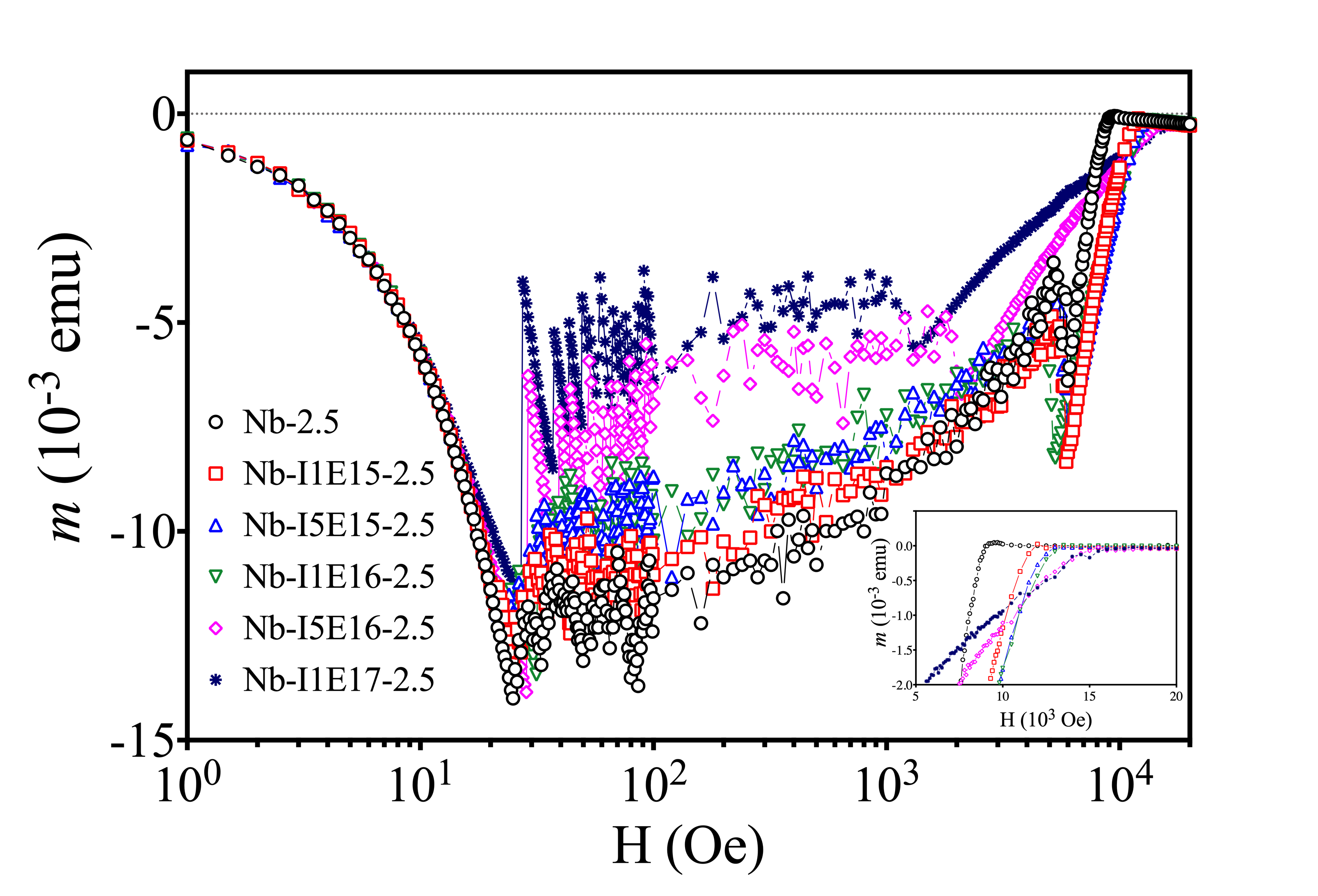}
\caption{
Magnetic moment as a function of applied magnetic field for the pristine Nb-2.5 film and the corresponding samples irradiated with Ar$^{+}$ at different fluences, measured after zero-field cooling to 2.0~K. The horizontal axis is plotted on a logarithmic scale. The curves are initially smooth, then exhibit successive discontinuities associated with thermomagnetic instabilities, and recover continuous behavior above the upper threshold field before the transition to the normal state at high fields. Inset: enlarged view of the high-field region of the \(m(H)\) curves, where the samples transition to the normal state at constant temperature.
}
\label{Fig7}
\end{figure}

Figure~\ref{Fig7} shows the magnetic moment as a function of applied field for the pristine reference sample Nb-2.5 and for samples irradiated with different Ar$^{+}$ fluences. Before each measurement, the samples were zero-field cooled from above \(T_c\) to 2.0~K. The Nb-2.5 film exhibits the typical response of a superconducting Nb film, with abrupt discontinuities appearing between the smooth virgin branch and the final portion of the curve, thus evidencing thermomagnetic instabilities, as discussed above.

The irradiated samples also exhibit field intervals in which thermomagnetic instabilities are present. Although only minor deviations from the pristine behavior are observed at low fluences, the effect of irradiation becomes pronounced for Nb-I5E16-2.5 and Nb-I1E17-2.5. At low fields, these samples display larger discontinuities, indicating larger flux avalanches and suggesting stronger pinning. At higher fields, the curves deviate from that of the pristine film and exhibit a reduced diamagnetic response. At the same time, the instability-termination field shifts to lower values with increasing fluence, thereby extending the magnetic-field range over which the samples remain stable.

The reduced diamagnetic response at high fields suggests a corresponding decrease in the shielding-current density. Within the thermomagnetic-instability framework, such a reduction favors reentrant stability. Yurchenko \textit{et al.} demonstrated that a monotonically decreasing $j_c(B)$ can produce an upper threshold field above which avalanche activity ceases because the critical flux-penetration depth required for the instability increases as $j_c$ decreases~\cite{yurchenko_reentrant_2007}. The progressive reduction in the upper threshold field observed for the irradiated samples in Figs.~7 and~8 is therefore consistent with these films approaching the thermomagnetically stable regime as the irradiation fluence increases. The reduced diamagnetic response, however, is not accompanied by a decrease in the upper critical field $H_{c2}$, at which the sample enters the normal state and $m(H)$ approaches zero. Instead, $H_{c2}$ increases with ion fluence in the irradiated samples~\cite{christen_formation_1987}, as shown in the inset of Fig.~\ref{Fig7}.

\begin{figure}[t!]
\centering\includegraphics[width=1.0\linewidth]{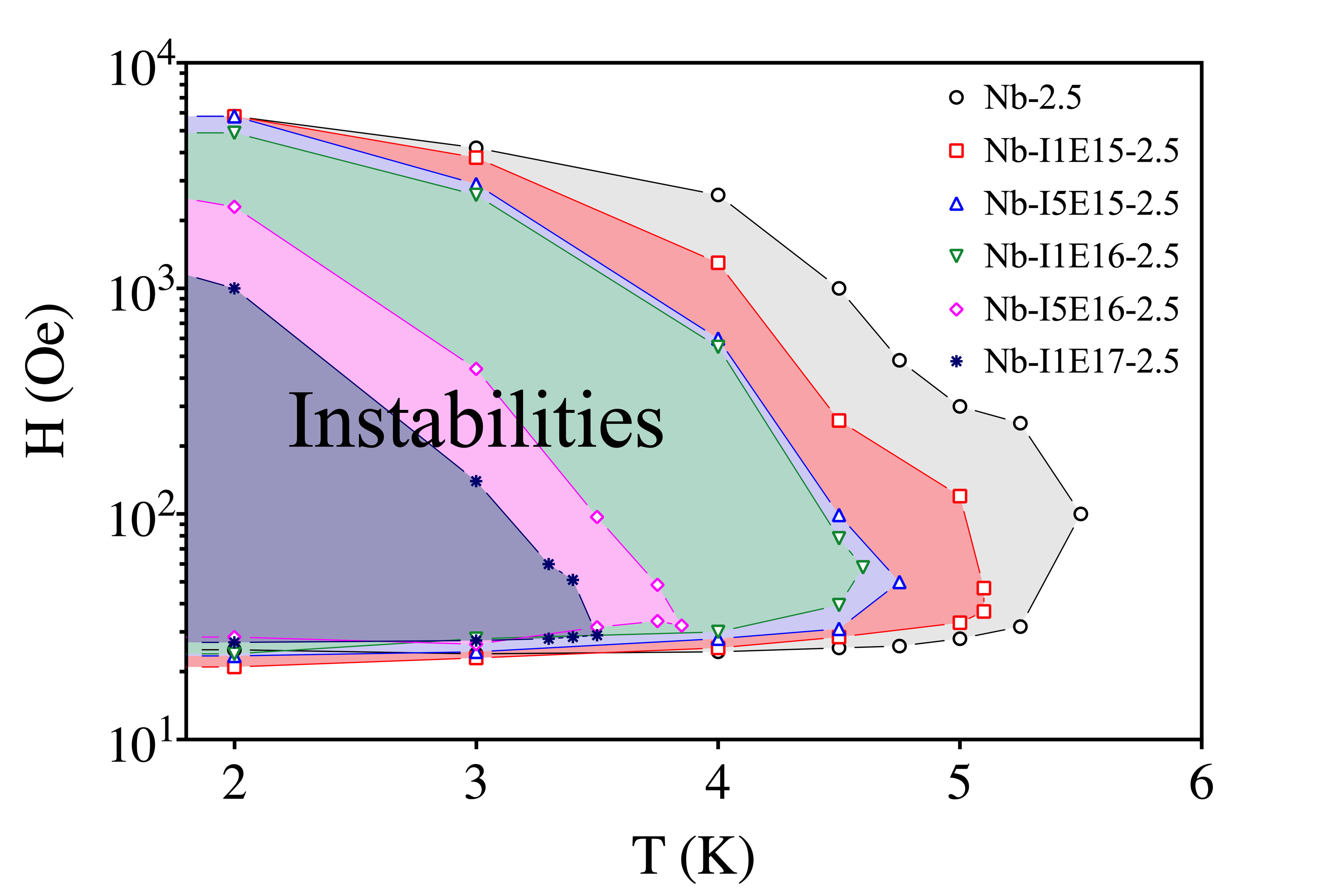}
\caption{
Field-temperature (\(H\)-\(T\)) diagram for the pristine Nb-2.5 film and the corresponding samples irradiated with Ar$^{+}$ at different fluences. The applied magnetic field is shown on a logarithmic scale. The region enclosed by the lower and upper threshold fields defines the field and temperature range over which thermomagnetic instabilities occur. The upper threshold field decreases significantly with increasing ion fluence, leading to a progressive reduction of the instability region.
}
\label{Fig8}
\end{figure}

Figure~\ref{Fig8} summarizes the boundaries of the thermomagnetic-instability regime in the \(H\)-\(T\) plane for the irradiated samples. 
The outermost contour corresponds to the instability region of the pristine film. 
The evolution of the upper boundary is particularly clear, since the area enclosed by the instability region decreases with increasing ion fluence. 
Ar$^{+}$ irradiation therefore extends the field-temperature range over which the film remains stable.

\subsection{Combined Effects}

Normal-metal layers and Ar$^{+}$ irradiation stabilize Nb films in complementary field regimes. Whereas the metallic overlayer is more effective in suppressing thermomagnetic instabilities at low fields, ion irradiation provides stronger stabilization at higher fields. Their combination therefore offers a route toward complete stabilization over the entire field range.

\begin{figure}[t!]
\centering\includegraphics[width=1.0\linewidth]{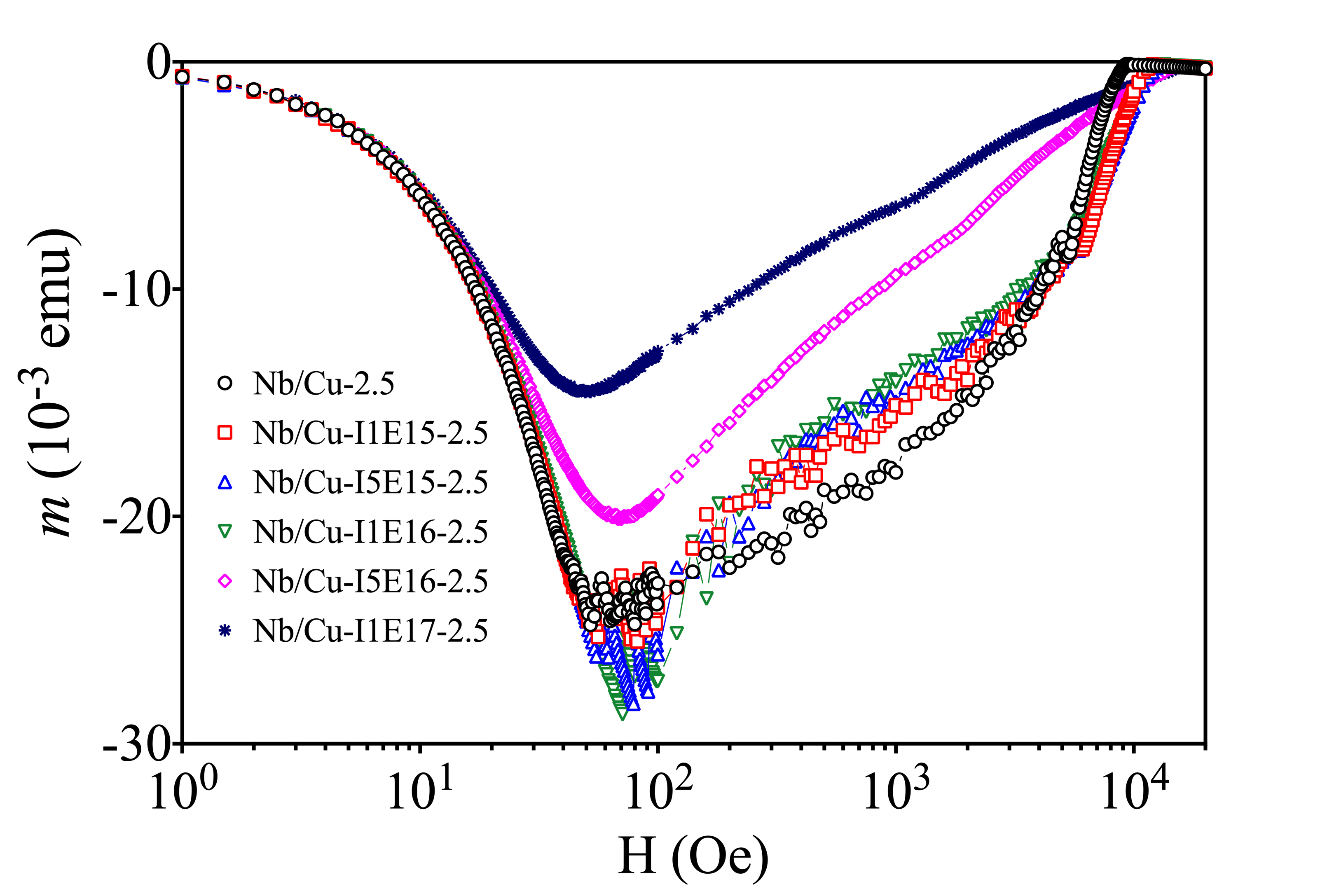}
\caption{
Magnetic moment as a function of applied magnetic field for Nb films irradiated with Ar$^{+}$ at different fluences and coated with a 1~\(\mu\)m-thick Cu layer, measured after zero-field cooling to 2.0~K. The horizontal axis is plotted on a logarithmic scale. The curves exhibit discontinuities associated with thermomagnetic instabilities at lower irradiation fluences, whereas the Nb/Cu-I5E16-2.5 and Nb/Cu-I1E17-2.5 samples remain smooth over the entire field range, indicating complete stabilization.
}
\label{Fig9}
\end{figure}

This combined effect is demonstrated in Fig.~\ref{Fig9}, where a 1~$\mu$m-thick Cu layer was deposited on samples previously subjected to ion irradiation. Samples irradiated at low fluences (Nb/Cu-I1E15-2.5, Nb/Cu-I5E15-2.5, and Nb/Cu-I1E16-2.5) still exhibit thermomagnetic instabilities despite the presence of the adjacent Cu layer. By contrast, samples irradiated at higher fluences (Nb/Cu-I5E16-2.5 and Nb/Cu-I1E17-2.5) exhibit fully smooth magnetization curves. These results show that thermomagnetic instabilities can be completely suppressed by combining Ar$^{+}$ irradiation at an appropriate fluence with a sufficiently thick Cu overlayer.

\begin{figure}[t!]
\centering\includegraphics[width=1.0\linewidth]{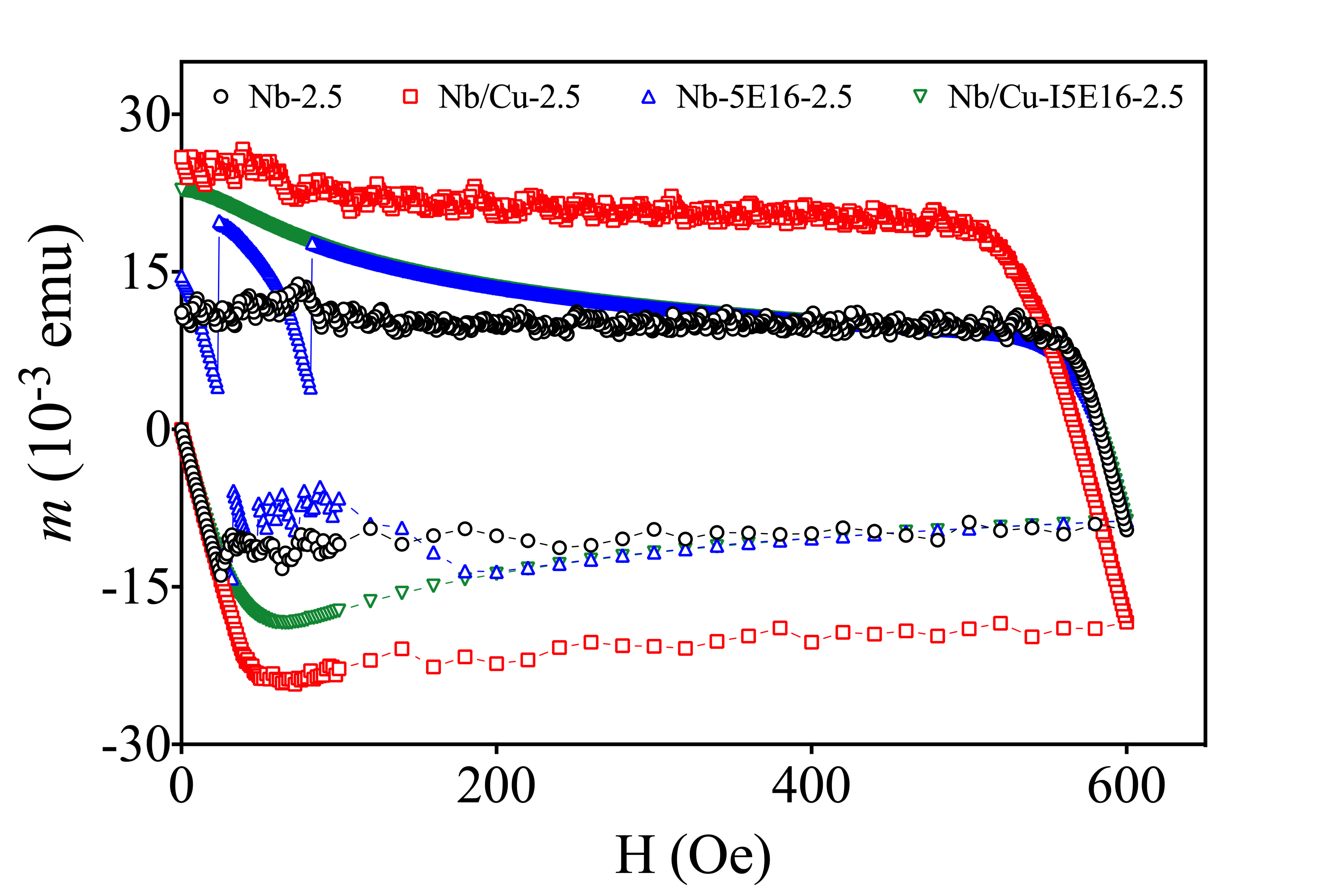}
\caption{
Magnetic moment as a function of applied magnetic field at 3.5~K for the Nb-2.5, Nb/Cu-2.5, Nb-I5E16-2.5, and Nb/Cu-I5E16-2.5 samples. The curves correspond to hysteresis half-loops. The pristine Nb-2.5 sample exhibits flux avalanches over the entire field range shown. In Nb/Cu-2.5, avalanche activity remains present, but the diamagnetic signal is enhanced. The irradiated Nb-I5E16-2.5 sample exhibits large avalanches at low fields and a nearly smooth descending branch, except for two pronounced events. The combined sample, Nb/Cu-I5E16-2.5, shows a fully smooth magnetization curve, indicating complete suppression of thermomagnetic instabilities.
}
\label{Fig10}
\end{figure}

To provide an overall view of the stabilization process, Fig.~\ref{Fig10} compares the hysteresis half-cycles measured at 3.5~K in applied fields up to 600~Oe. 
The pristine sample, Nb-2.5, exhibits an initial smooth regime up to about 30~Oe, above which thermomagnetic instabilities develop and persist along both the ascending and descending branches. For the film coated only with a 1~$\mu$m-thick Cu layer, Nb/Cu-2.5, this smooth regime extends to about 50~Oe. 
In addition, the diamagnetic response is enhanced and the magnetization jumps become smaller, including those along the descending branch. 

When the film is irradiated only with Ar$^{+}$ at a fluence of \(5\times10^{16}\)~ions/cm\(^2\), corresponding to sample Nb-I5E16-2.5, the lower threshold field remains close to 30~Oe.
However, the jumps at low fields are large, whereas at higher fields the diamagnetic signal reaches values comparable to those of the pristine sample. 
An important effect of irradiation is the nearly complete suppression of instabilities along the descending branch, except for the two largest events. The sample Nb/Cu-I5E16-2.5 combines both a metallic overlayer and Ar$^{+}$ irradiation. 
In this case, the magnetization remains smooth along both the ascending and descending branches, while at higher fields its diamagnetic magnitude is similar to that of the pristine sample. 
The overlap of the initial virgin curves reflects the fact that the Nb films used for the pristine and modified samples were obtained from the same batch.
Remarkably, this combination yields a sample that is completely free of instabilities.

\begin{figure}[t!]
\centering\includegraphics[width=1.0\linewidth]{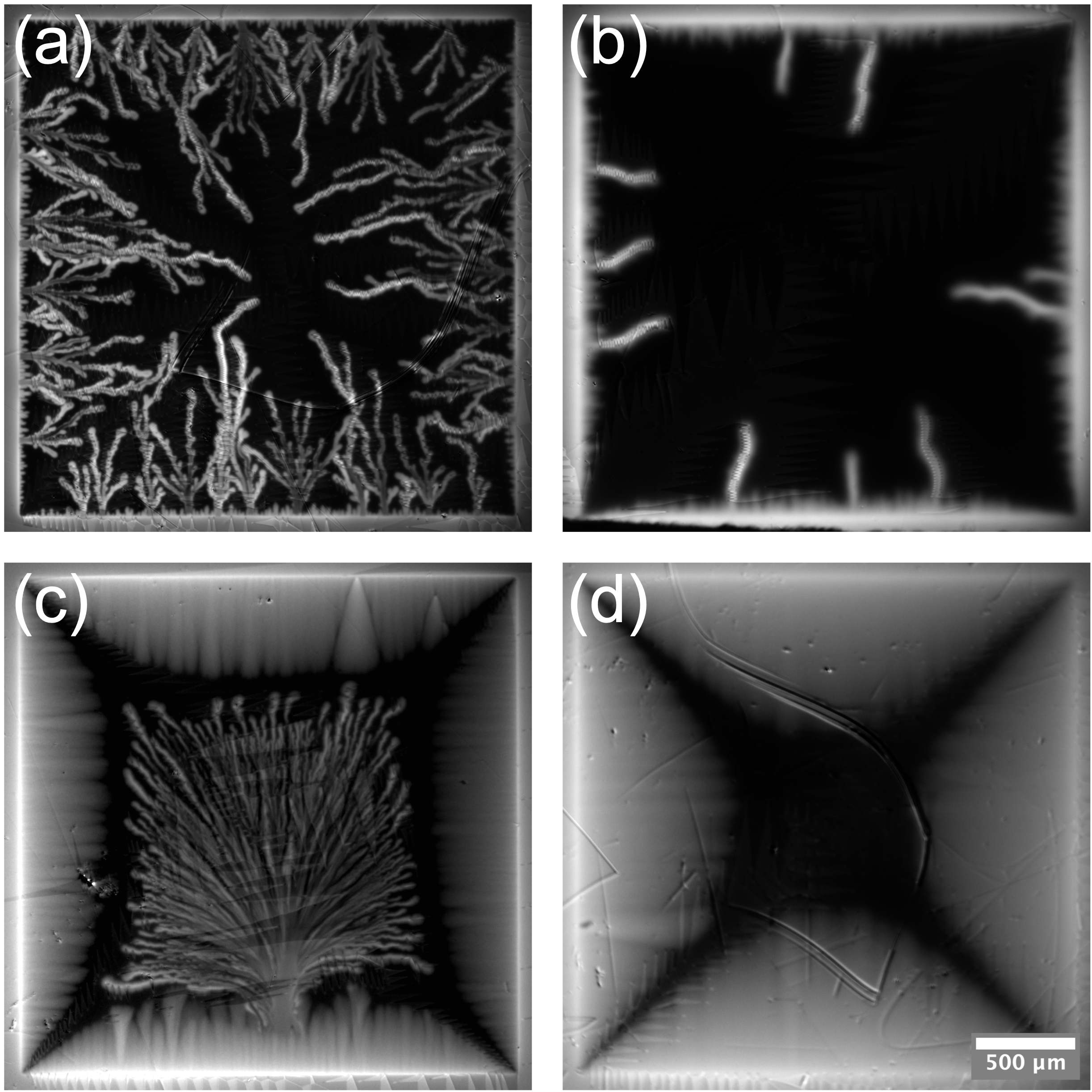}
\caption{
Magneto-optical images of flux penetration acquired at 3.5~K and 60~Oe after zero-field cooling, illustrating successive stages of the stabilization process. In panel~(a), the Nb-2.5 sample exhibits long dendritic avalanches. In panel~(b), the avalanches in Nb/Cu-2.5 are shorter, and the larger dark region indicates enhanced screening of the applied field. In panel~(c), a single large dendritic avalanche in Nb-I5E16-2.5 extends over a substantial portion of the sample area. In panel~(d), no avalanches are observed in Nb/Cu-I5E16-2.5, and flux penetrates smoothly from the sample edges, consistent with stable critical-state behavior.
}
\label{Fig11}
\end{figure}

The magneto-optical images in Fig.~\ref{Fig11} show the flux penetration at 3.5~K and 60~Oe for the samples whose magnetization curves are presented in Fig.~\ref{Fig10}. Panel (a) shows the flux pattern accumulated in Nb-2.5 during the initial field increase. The multiple long branches correspond to the successive jumps observed in the magnetization curve. Under the same conditions, panel (b) shows that flux penetration in Nb/Cu-2.5 occurs through fewer and smaller fingers. A larger dark region is also visible, reflecting stronger magnetic-field screening and being consistent with the enhanced diamagnetic response. 
Panel (c) shows the first avalanche nucleated in Nb-I5E16-2.5, along with additional smooth flux penetration from the edge.
The dendrite forms a highly branched structure originating from the lower edge and extending over most of the sample area, in agreement with the first prominent jump in the magnetization curve of Fig.~\ref{Fig10}.
Finally, panel (d) shows that no flux avalanches occur in Nb/Cu-I5E16-2.5 under these conditions. Instead, flux penetrates smoothly from all sides, consistent with the continuous magnetization curve in Fig.~\ref{Fig10}. 
These results confirm that the combined treatment restores the ability of the film to screen the applied field without being disrupted by flux instabilities.

\section{Summary}

Thermomagnetic instabilities in Nb superconducting films were investigated using two distinct strategies: the addition of normal-metal overlayers and Ar$^{+}$ irradiation. The experiments show that these modifications stabilize the films in complementary field regimes. The metallic layers suppress the onset of flux avalanches at low applied fields, with the strongest effect observed for Cu, consistent with its high electrical conductivity and the correspondingly large value of the suppression parameter \(S\). By contrast, ion irradiation reduces the high-field portion of the instability region by shifting the upper threshold for avalanche activity to lower fields.

When combined in the same sample, these two partial stabilization effects lead to complete suppression of thermomagnetic instabilities. In particular, Nb/Cu-I5E16-2.5 exhibits smooth magnetization curves and avalanche-free flux penetration, as confirmed by both magnetometry and magneto-optical imaging. This combined treatment restores stable critical-state flux penetration and preserves the ability of the superconducting film to screen the applied magnetic field without intermittent flux-jump activity.

These results establish a practical route for stabilizing Nb thin films against thermomagnetic avalanches through post-growth modification and hybrid design. Beyond their relevance to the fundamental understanding of flux instabilities in superconducting films, the present findings are particularly significant for superconducting thin-film technologies, in which uncontrolled flux motion can compromise the performance, stability, and precision of sensitive electronic devices.

\begin{acknowledgments}
The samples were grown at the Laboratório de Conformação Nanométrica (LCN-IF-UFRGS), lithography was carried out at the Laboratório de Micro e Nano Fabricação (MNF/LNNano/CNPEM, Proposals~16254 and 24642), and structural characterization was performed at the Laboratório de Corte e Orientação de Cristais (LCOC/LNLS/CNPEM, Proposal~20262659). This work was supported by the São Paulo Research Foundation (FAPESP, Grants~2017/24786-4, 2021/08781-8, and 2025/23429-0) and the Conselho Nacional de Desenvolvimento Científico e Tecnológico (CNPq, Grants~151569/2012-6, 302586/2018-0, 434797/2018-9, 431974/2018-7, and 306894/2022-0). We also acknowledge support from the INCT Advanced Quantum Materials project, funded by the Brazilian agencies CNPq (Proc.~408766/2024-7), FAPESP (Proc.~2025/27091-3), and CAPES.
\end{acknowledgments}

\bibliography{apssamp}

\end{document}